\documentclass[journal,final,letterpaper]{IEEEtranTCOM}

\usepackage[hang,flushmargin]{footmisc}

\usepackage{amssymb, graphicx, dsfont, flushend}
\usepackage[cmex10]{amsmath}
\usepackage{footmisc}	
\usepackage{subcaption}
\usepackage{textcomp}
\usepackage{algorithm}
\usepackage{algpseudocode}
\usepackage{ulem,color}

\newcommand{\newt}[1]{#1}

\begin{document}

\title{
Spatial Reuse in Dense Wireless Areas: A Cross-layer Optimization Approach via ADMM
}
\author{Haleh~Tabrizi,~\IEEEmembership{Member,~IEEE,}
				Borja~Peleato,~\IEEEmembership{Member,~IEEE,}
				Golnaz~Farhadi,~\IEEEmembership{Member,~IEEE,}
         John~M.~Cioffi,~\IEEEmembership{Fellow,~IEEE,}
        and~Ghadah~Aldabbagh,~\IEEEmembership{Member,~IEEE} 
\thanks{H. Tabrizi is with the Department of Electrical Engineering, Stanford University, Stanford, CA, 94305 USA, e-mail: (htabrizi@stanford.edu)}
\thanks{B. Peleato is with the Department of Electrical and Computer Engineering, Purdue University, West Lafayette, IN, 47907 USA, e-mail: (bpeleato@purdue.edu).}
\thanks{J. Cioffi is with the Department of Electrical Engineering, Stanford University, Stanford, CA, 94305 USA, and Department of Computer Science, King Abdulaziz University, Jeddah, Saudi Arabia, e-mail: (cioffi@stanford.edu).}
\thanks{G. Farhadi is with Fujitsu Labs of America, Sunnyvale, CA, 94085 USA, e-mail: (gfarhadi@us.fujitsu.com).}
\thanks{G. Aldabbagh is with the Department of Computer Science, KAU, Jeddah, Saudi Arabia, e-mail: (galdabbagh@kau.edu.sa)}
\thanks{This paper has been submitted in part to IEEE global communications conference, Atlanta, GA, December 2013. }
\thanks{\copyright2015 IEEE.  Personal use of this material is permitted.  Permission from IEEE must be obtained for all other uses, in any current or future media, including reprinting/republishing this material for advertising or promotional purposes, creating new collective works, for resale or redistribution to servers or lists, or reuse of any copyrighted component of this work in other works. DOI: 10.1109/TWC.2015.2464373}}

\markboth{IEEE Transactions on Wireless Communications}
{Published paper}

\maketitle

\begin{abstract}
This paper introduces an efficient method for communication resource use in dense wireless areas where all nodes must communicate with a common destination node. The proposed method groups nodes based on their \newt{distance from the destination} and creates a structured multi-hop configuration in which each group can relay its neighbor's data. \newt{The large number of active radio nodes and the common direction of communication toward a single destination are exploited to reuse the limited spectrum resources in spatially separated groups}. Spectrum allocation constraints among groups are then embedded in a joint routing and resource allocation framework to optimize the route and amount of resources allocated to each node. \newt{The solution to this problem uses coordination among the lower-layers of the wireless-network protocol stack to outperform conventional approaches where these layers are decoupled. Furthermore, the structure of this problem is exploited to obtain} a semi-distributed optimization algorithm based on the alternating direction method of multipliers (ADMM) where each node can optimize its resources independently based on local channel information.
\end{abstract}


\section{Introduction}

With the remarkable growth of wireless technology and advancement of mobile devices, the demand for more efficient use of limited radio resources is increasing rapidly. Optimizing performance through sophisticated physical-layer techniques is no longer enough, so studies have considered coordinating multiple users through scheduling and admission control, as well as joint operation of the physical and protocol layers in wireless networks. Lin et al. summarize such cross-layer optimization techniques in \cite{lin06}.

This paper studies efficient resource use in dense wireless areas where all nodes communicate with a single destination node. Dense wireless areas are locations populated with many radio devices that must communicate over a shared spectrum. Some examples include concert halls and stadiums populated with user handheld devices, which must all communicate with a common access point or base-station (BS). Home Area Networks (HAN) are another example of dense wireless areas, particularly when such residential local area networks may eventually connect hundreds of digital devices within a home in the so-called ``Internet of things''.

\begin{figure}[t]
\centering
\includegraphics[scale=0.25]{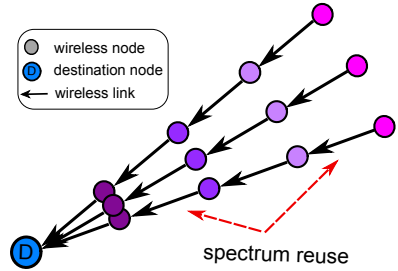}
\caption{\small Multiple hops and reuse between hops.}\label{fig:hops} 
\end{figure}

Figure~\ref{fig:hops} suggests a configuration that can increase spectrum efficiency by relaying data over multiple hops rather than direct communication. The node configuration motivates a hierarchical multi-hop architecture, where spectrum can be reused among the hops as well as among cells \cite[Ch. 15]{goldsmith}. Furthermore, the presence of a large number of nodes in a dense wireless area offers many possible routes that the proposed multi-hop configuration can optimize. This approach does not require additional infrastructure, but the nodes act as relays for their neighbor's data.

In this scenario all nodes communicate with a fixed destination node, hence the reuse pattern and flow of data can be structured. The nodes that are in close proximity of each other are grouped together to form an FDMA (frequency division multiple access) system where each node operates on different spectra. Each such group of nodes can relay data for neighboring groups\footnote{A configurable MAC (e.g. platform proposed in \cite{neu05}) may be used to enable a different channel access method (random access, FDMA, etc.) for different nodes depending on their congestion level along the path.} along an optimal routing path determined by the capacity of the links between nodes. The capacity of each link is determined by the power and bandwidth allocated to that link, which in turn depend on the total spectrum allocated to each group. Finding the optimal network configuration therefore requires a joint optimization considering all constraints.

 Joint optimization and coordination of the different network layers is not a new concept. It has been used in many different applications in wireless and wireline networks \cite{san03}, \cite{vin05}. Simultaneous routing and resource allocation methods for orthogonal multiple access systems have been proposed by \cite{boyd04} (FDMA and TDMA) and \cite{boyd03} (CDMA). Both approaches jointly optimize network-layer routing and physical-layer resource allocation through a convex formulation of the problem. A joint scheduling, routing, and resource allocation for multi-hop cellular networks with relay stations is proposed by \cite{tim11}. In this case, the general problem is non-convex, and hence an iterative algorithm is used to coordinate the layers. Beyond resource allocation and relay selection, \cite{yu07} jointly optimizes the relay strategy that should be used between each source and destination pair based on a cross-layer design.

This paper addresses an extended version of the simultaneous routing and resource allocation (SRRA) framework proposed by \cite{boyd04}, including additional constraints on the total available spectrum resources and spectrum reuse among the predetermined user groups. The joint optimization of routing and global resource allocation in a dense area is a non-convex problem that \newt{can only be solved exactly through exhaustive search methods, such as branch and bound. Unfortunately, these methods are often slow and have exponential worst-case performance \cite{BB}. Hence, we propose simplifying the problem into a convex form that can be solved using ADMM (Alternating Direction Method of Multipliers), which has $O(1/k)$ convergence rate \cite{boyd11}.}

 The predetermined user groups are selected based on each node's relative distance to the destination and a subset of all possible links connecting nodes from one group to a neighboring group are selected as candidate data flow links in the optimization problem. The goal is to achieve jointly optimized routing, \emph{global} spectrum assignment, and local resource allocation for the selected links through convex optimization.

  Distributed implementations of network utility maximization problems often lead to simpler subproblems, each optimizing a subset of the decision variables based on local information. \newt{Palomar et al. \cite{mung07} propose a systematic framework to exploit decomposition structures that lead to different distributed algorithms. A very popular approach is using dual decomposition to generate a highly distributed optimization problem \cite{boyd04}, but some studies seek faster convergence using Newton's method \cite{jia12}}.  Certainly, there is a tradeoff between the amount of message-passing among computing entities, convergence speed, and computational complexity of each distributed algorithm. This paper exploits the structure in the proposed joint optimization framework to generate two different distributed algorithms: The first algorithm (centralized) separates the problem into two subproblems, allowing decoupled optimization of network-layer and physical-layer variables. The second algorithm (semi-distributed) employs all active nodes as computing resources that perform parallel optimization based on local channel information.

 The proposed decompositions use the alternating direction method of multipliers (ADMM) \cite{boyd11} algorithm, which is more numerically stable than conventional decomposition methods. For example, ADMM does not require strict convexity of the objective, but the dual decomposition method proposed in \cite{boyd04} does. ADMM has been widely employed in producing distributed algorithms. Some recent examples include multicell coordinated beamforming \cite{shen12} and distributed model predictive control \cite{mot12}.


 The main contributions of this paper are: 1) devising a hierarchical structure of nodes that allows resource reuse among spatially separated nodes, 2) integrating global resource allocation and reuse in a cross-layer optimization framework of routing and resource allocation, and 3) developing a semi-distributed optimization algorithm based on ADMM where each node optimizes the resources based on its local channel information.

The rest of this paper is organized as follows. \newt{Section~\ref{sec:systemModel} describes the system model, consisting of the network topology and the physical layer constraints in a dense wireless area. Section~\ref{sec:algorithm} formulates the cross-layer optimization problem for the previous model in a convex form. Section~\ref{sec:distributed} proposes a centralized and a distributed algorithm to solve this problem, both based on ADMM. It also provides a numerical example to illustrate the performance and convergence of the algorithm for a small network. Finally, Section~\ref{sec:performance} presents more detailed simulations evaluating the performance of the algorithm on larger networks and Section~\ref{sec:summary} concludes and summarizes the paper}.

\section{Problem Formulation}\label{sec:systemModel}
The system model and the proposed hierarchical node configuration dictate a set of system constraints that need to be satisfied while optimizing network performance. This section formulates and motivates these constraints.

\subsection{Network Flow Model} \label{subsec:networkFlow}
We consider $N$ active users/nodes randomly distributed in a given area that need to communicate simultaneously with a common destination node, such as a cellular base station (BS). This paper investigates uplink transmission, but downlink communication can be formulated in a similar manner with minor modifications. A standard directed graph is used to represent the network topology. The graph is assumed to be connected, \textit{i.e.}, there is a route between every node and the BS. The nodes are labeled as $n \in \mathcal{N} = \left\{1, 2, ..., N\right\}$, with node $1$ representing the BS.  All nodes can transmit, receive and relay data over the existing links. Transmission and reception occur on disjoint spectra.  A link exists between two nodes $i$ and $j$ if direct communication among the two is possible. With a total of $L$ links in the network, denote the set of all links by $\mathcal{L} = \left\{1,2,...,L\right\}$, where each link is identified with an integer value between $1$ and $L$.

\begin{figure}[t]
\centering
\includegraphics[scale = 0.2]{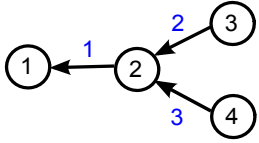}
\caption{\small Simple node and link example.}\label{fig:linkIncidance}
\end{figure}

The network topology is represented by a link-node incidence matrix $A \in \mathbb{R}^{N \times L}$, whose entries determine a link's source and destination nodes by $+1$ and $-1$, respectively. For example, for the configuration shown in Figure~\ref{fig:linkIncidance}, where $N=4$ and $L=3$, the link-incidence matrix is given by: $A = \Bigg(\begin{smallmatrix}
-1 &  0  & 0 \\
+1 & -1  & -1 \\
0  & +1  & 0 \\
0  &  0  & +1 \end{smallmatrix} \Bigg)$. Each column of $A$ corresponds to a link and each row corresponds to a node. The first column indicates that link $1$ goes from node $2$ to node $1$. The set of outgoing links from node $n$ are denoted by $\mathcal{O}\left(n\right)$ and the set of incoming links are denoted by $\mathcal{I}\left(n\right)$. 

 A multi-commodity flow model \cite{berts} that satisfies flow conservation at each node is considered. For the uplink case considered here, $x \in \mathbb{R}^{L}$ identifies the flow on each link (data rate) that is destined to node $1$ (BS). Let $A_{nl}$ denote the element on the $n$-th row and $l$-th column of $A$, and let $r \in \mathbb{R}^{N}$ be a vector of data traffic injected in the network at each node. Then $Ax = r$, where $r_n=\sum_l{A_{nl}x_l}$, denotes the amount of data traffic injected into the network at node $n$. Positive components will indicate flow sources, while negative ones will indicate flow sinks. For the example in Figure \ref{fig:linkIncidance}, the flow conservation law indicates that $r_1 = - x_1$, $r_2 = x_1 - x_2 - x_3$, $r_3 = x_2$, and $r_4 = x_3$.
Furthermore, the amount of flow on each link ($x_l$) is constrained by that link's physical layer capacity $c_l$:
\[
x_l \leq c_l,\quad \forall l \in \mathcal{L}.
\]

\subsection{Physical Layer Model} \label{subsec:physicalLayer}

An FDMA channel access method is considered, where each node's outgoing links are assigned disjoint frequency bands. The received signal is corrupted by additive white Gaussian noise with power spectral density $\sigma_l$ and the link capacity is given by a concave and monotone increasing function of its allocated bandwidth $w_l$ and power $p_l$, as follows:
\[
c_l = w_l\log\left( 1+\frac{p_lq_l}{w_l\sigma_l}\right), \quad \forall l \in \mathcal{L}.
\]
Parameter $q_l$ is the channel gain from the source to the destination of link $l$. The total amount of power allocated to a node $n$ is constrained to be smaller than a pre-fixed value $P_{\max, n}$, hence
\begin{equation*}
\sum_{l\in \mathcal{O}(n)}{p_l\leq P_{\max, n}},\quad \forall n \in \mathcal{N}.
\end{equation*}
The total amount of bandwidth allocated to node $n$ is denoted by
\begin{equation*}
\sum_{l \in \mathcal{O}(n)}{w_l} = v_{n},\quad \forall n \in \mathcal{N}. \label{perNodeBW}
\end{equation*}

If necessary, after obtaining the continuous bandwidth variables $w_l$, they can be quantized based on the underlying system modulation approach. For example, if the underlying system is an OFDMA system, the bandwidth values can be rounded down to an integer number of subcarrier bandwidths and if the underlying system is an LTE system, it can be rounded to the closest integer number of resource blocks.

\subsection{Group Assignment} \label{subsec:groupAssignment}
 The objective of the proposed multi-hop configuration is to reuse as much of spectrum as possible. Hence, direct links between far nodes and the BS are divided into multiple hops that operate on different spectrum bands, allowing reuse (Figure~\ref{fig:hops}). To this end, nodes are grouped based on their distance to the BS as shown in Figure~\ref{fig:links}: nodes located between radial distances $d_{g-1}$ and $d_{g}$ from the BS belong in group $\mathcal{G}_g$, and all nodes within distance $d_1$ of the destination belong to group $\mathcal{G}_1$. The coverage area is thus partitioned into $M$ disjoint sections such that $\mathcal{G}_g \bigcap \mathcal{G}_{g'} = \emptyset,\quad \mathcal{G}_g\neq\emptyset,\quad \forall g,g' $, and each node belongs to a group: $\bigcup_{g=1}^M \mathcal{G}_g = \mathcal{N}$.



In general, the closer a link to the BS, the larger the flow it will be carrying. The total flow on the links that terminate at the BS is the largest and the flow decreases as the links move away from the BS as shown in Figure~\ref{fig:links}. Let $f$ denote the frequency reuse factor, such that every $(f+1)$-th group uses the same spectrum set ($f>2$ to avoid the hidden node problem). As such, the maximum available spectrum $W_{\max}$ is divided among groups $1$ through $f$:
\[
\sum_{g=1}^{f}{W_{g}}  = W_{\max},
\]
where $W_g$ denotes the amount of bandwidth allocated to group $g$. \newt{The rest of the groups ($f+1$ through $M$)} reuse the spectrum allocated to groups $1$ through $f$:
\[
W_{g} = W_{ \left( (g-1)\bmod{f} \right) +1}, \; \forall g > f.
\]
For example, if $f = 3$, the total bandwidth is divided between groups $1, 2$ and $3$, while groups $4, 7,10,\ldots$ use the same spectrum as group $1$; groups $5, 8,11,\ldots$ use the same spectrum as group $2$; and \newt{ groups $6,9,12,\ldots$ use the same as group $3$.}


\subsection{Routing Link Assignments} \label{subsec:linkAssignment}
In order to optimize the routing from each node to the BS, an initial set of possible links should be determined. This subsection describes a method, similar to that in \cite{san10}, for selecting this initial pool of links \newt{(or refining a pre-defined set of links if desired). The method assumes that the BS knows the location of each node and uses relative distances between them to select the initial pool of links. If this assumption does not hold, the initial pool of links could include all those between nodes in consecutive groups. Regardless of how the initial pool of links is selected, resource and flow allocation are based on channel information.}

\begin{figure}[t]
\centering
\includegraphics[scale=0.3]{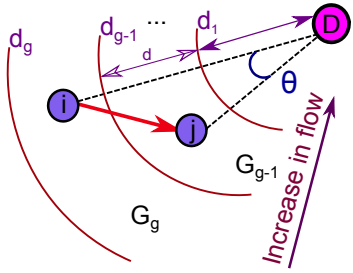}
\caption{\small Link assignment conditions.}\label{fig:links}
\end{figure}

\newt{The proposed method will include a link from node $i$ to node $j$ in the initial pool} if:
\begin{enumerate}
	\item $j \in \mathcal{G}_{g-1}$, where $i \in \mathcal{G}_g$.
	\item the angular distance between nodes $i$ and $j$ is less than a pre-fixed $\theta$, as shown in Figure~\ref{fig:links}.
	\item the distance between nodes $i$ and $j$ is less than a pre-fixed $d_{th}$.
\end{enumerate}
The first condition enforces a single hop between consecutive groups to allow spectrum reuse. If $i \in \mathcal{G}_g$ and $j \in \mathcal{G}_h$ with $h < g-1$, then the link from $i$ to $j$ traverses more than one group (spectrum region) and defeats the purpose of reuse. On the other hand, if $h = g$ then $i$ and $j$ belong to the same group creating an extra hop within the group, and if $h>g$, then $j$ is farther away from the BS than $i$ already is. In both cases, $j$ has no relaying benefit for $i$.

The second and third conditions ensure that the relay node $j$ is not too far away from the source node $i$. Parameters $d_{th}$ and $\theta$ control the number of outgoing links from a node and allow selecting a reasonable \newt{(non-empty)} set of possible routes.

The maximum transmission power per unit bandwidth per link is limited to avoid interference between groups operating on the same spectrum. This maximum transmission power depends on the radial distances $d_g$, $g = \left\{1,2,...,M\right\}$ defining the groups and on the frequency reuse factor $f$. \newt{If every group has the same radial distance $d$, the minimum possible distance} between the transmitter of one link and a receiver suffering its interference is equal to $d_{int} = (f-2) \times d$. The received power per unit bandwidth has to be less than a fraction $\alpha$ of the noise power for interference to be negligible, hence the constraint on the transmission power is:
\[
p_l \leq w_l \frac{\alpha N_0}{K(\frac{l_0}{d_{int} })^a},
\]
where $a$ is the pathloss exponent, $l_0$ is the reference distance\footnote{This is a conservative bound, using the shortest possible distance between interfering nodes. Results could be improved using the actual distance for each node, if static and known.}, \newt{$N_0$ is the noise power spectral density at the receiver ($\sigma_l$),} and $K$ is the attenuation factor based on the simplified pathloss model \cite{goldsmith}. This work denotes the bandwidth coefficient by
\[
\gamma=\alpha N_0\left( K\left( \frac{l_0}{d_{int}} \right)^a \right)^{-1},
\]
such that the per link transmission power constraint becomes $p_l \leq w_l\gamma$.

\section{Joint Routing, Global and Local Resource Allocation}\label{sec:algorithm}

Based on the network conditions, such as node density and locations, the user groups and possible routing links are determined. Given these, the routing and resource allocation problem can be formulated as a convex optimization problem with the objective of maximizing a concave utility function. To maintain fairness among all users, the utility function was chosen to be the minimum rate among all users ($n \in \mathcal{N}\setminus \left\{1\right\}$). Only uplink communication is considered here, so the set of all transmitters (all nodes other than the BS) is $\mathcal{\tilde{N}} = \mathcal{N} \setminus \left\{1\right\}$.

For each link $l \in \left\{1,..L\right\}$ in the initial pool, we define the system variables $p_l$ and $w_l$, respectively denoting the amount of power and spectrum allocated to the link, and $x_l$, denoting the amount of flow on the link destined to the BS. $W_g, \; \forall g \in \left\{1,...,M\right\}$ will denote the amount of spectrum allocated to each group. The problem then becomes:
\begin{subequations}
\begin{align} \label{jointOpt}
	&\underset {p, w, x,(r_n,W_g)}{\text{maximize}}
	& &\min_{n \in \mathcal{\tilde{N}} }{r_n} \tag{1}\\
	&\text{subject to}
	& &\sum_l{A_{nl}x_l} = r_n, \; \forall n \in \mathcal{N} \label{flowCons}\\
	&&&x_l\leq  w_l\log\left( 1+\frac{p_lq_l}{w_lN_0}\right), \; \forall l \in \mathcal{L} \label{flowCap} \\
	&&& p_l \leq w_l \gamma, \; \forall l \in \mathcal{L} \label{powerConstraint}\\
	&&&\sum_{l\in \mathcal{O}(n)}{p_l\leq P_{\max}}, \; \forall n \in \mathcal{\tilde{N}}   \label{PmaxConstraint} \\
	&&&\sum_{n \in \mathcal{G}_g}{\sum_{l \in \mathcal{O}(n)}{w_l} } \leq W_{g}, \; \forall g \in \left\{1,2,...,M\right\} \label{layergBW} \\
	&&&\sum_{g=1}^{f}{W_{g}}  = W_{\max} \label{bwConstraint}\\
	&&&W_{g} = W_{ \left( (g-1)\bmod{f} \right) +1}, \; \forall g > f\label{individualBwConstraints}\\
	&&& x_l \geq 0, \; p_l \geq 0, \; w_l \geq 0, \; \forall l \in \mathcal{L}. \label{positivity}
\end{align}
\end{subequations}

Constraints (\ref{flowCons}) through (\ref{powerConstraint}) are as explained in Section~\ref{sec:systemModel}. Constraint~(\ref{PmaxConstraint}) is obtained by setting the maximum transmission power of all users equal to $P_{\max}$. 
Constraints (\ref{layergBW}) through (\ref{individualBwConstraints}) identify the spectrum reuse constraints introduced in Section \ref{sec:systemModel}. The variables $W_g$ and $r_n$ are auxiliary variables introduced to make the problem formulation easier to understand.  The value for $r_n$ can be derived from $x$ using equations (1a). Similarly, it is easy to prove that constraint (1e) will be tight in the optimal solution, so the value for $W_g$ can also be uniquely derived from $w$. Consequently, both $r_n$ and $W_g$ appear between brackets in the list of optimization variables. This notation for auxiliary variables continues throughout the rest of the paper.

The constraints in (\ref{jointOpt}) define a convex set and the objective function is concave. Hence, the problem is convex and can be solved globally and efficiently by interior-point methods \cite{boyd}. However, there might be multiple solutions with the same objective value. Adding a small negative factor of power sum, $-\epsilon \sum{p_l}$, to the objective would guarantee that a solution with small power values is selected among the optimal set. This penalty term has negligible effect on the resulting \newt{minimum} rate value $r_n$.

The conventional direct-mode resource-allocation problem can be derived as a specific scenario of the multi-hop optimization problem (\ref{jointOpt}). In this formulation, there is only one group of nodes, and the only possible route from each node to the destination node is a direct link between the node and the destination. In this case, the problem can be rewritten as:
\begin{subequations}
\begin{align} \label{DM}
	&\underset {p, v,(r_n)}{\text{maximize}}
	& &\min_{n \in \mathcal{\tilde{N}} }{r_n} \tag{2}\\
	&\text{subject to}
	& & 0 \leq r_n\leq  v_n\log\left( 1+\frac{p_nq_n}{v_nN_0}\right), \; \forall n \in \mathcal{\tilde{N}} \nonumber \\
	&&&0 \leq p_n\leq P_{\max}, \; \forall n \in \mathcal{\tilde{N}}   \nonumber \\ 
	&&&\sum_{n \in \mathcal{\tilde{N}} }{v_n} = W_{\max}, \;  \nonumber \\ 
	&&&v_n \geq 0, \; \forall n \in \mathcal{\tilde{N}},  \nonumber
\end{align}
\end{subequations}
where $v_n$ and $p_n$ are equivalent to the original parameters $w_l$ and $p_l$, respectively, since there is only one outgoing link from each node.

\section{Distributed Optimization}\label{sec:distributed}

Dual decomposition is an old method \newt{for solving a convex optimization problem in a (semi-) decentralized manner}\footnote{\newt{Contrary to popular belief, dual decomposition does not provide a fully distributed solution. There still needs to be a central entity gathering information from all nodes.}} \cite{ned09}. Generally, if the objective function is separable in its variables, the problem can be split into smaller subproblems. These subproblems can then be solved iteratively in parallel based on dual ascent (or descent). However, this method requires some hard assumptions such as strict convexity and finiteness of the objective function; otherwise it can result in numerical instability.

Augmented Lagrangian methods such as the method of multipliers achieve better convergence results by adding a penalty parameter to the dual ascent objective; however, the penalty parameter is not separable when the original problem is. The alternating direction method of multipliers (ADMM), on the other hand,  is an algorithm that combines the \textit{decomposability} property of dual ascent with the robust convergence properties of the method of multipliers \cite{boyd11}.

The objective function in problem (\ref{jointOpt}) is not strictly convex so dual decomposition cannot be applied directly. A weighted quadratic regularization term would make the objective strictly convex, but finding the appropriate weights is a challenging problem. If the weights are too small, dual decomposition will still suffer from numerical stability, while excessive weights would involve an optimality cost.  ADMM is in general more numerically stable and faster in convergence than the conventional dual decomposition method \cite{shen12}. This paper thus proposes ADMM to implement distributed cross-layer optimization algorithms for spatial reuse.


\subsection{Disjoint Network-Layer and Physical-Layer Optimization} \label{2level}
One common method for developing efficient algorithms in cross-layer optimization problems is to decouple the network-layer and physical-layer problems \cite{boyd04}, \cite{mung07}. This method divides the problem into two simpler subproblems, the optimization of network-layer and physical-layer variables, which can be solved separately and updated iteratively by exchanging messages between the two layers. This section takes a similar approach in decoupling the problems through ADMM. At each iteration $k \in \mathbb{Z}$, the network layer and physical layer subproblems are solved in parallel and based on the results, the dual variables are updated.

The network-layer parameters $x$ and the physical-layer parameters $w$ and $p$ are coupled by constraint~(\ref{flowCap}) in the above centralized optimization problem. To decompose the problem, a new auxiliary variable $t \in \mathbb{R}^L$ and $L$ equality constraints $t_l = x_l, \; \forall l$ are introduced. ADMM decouples constraints $t_l = x_l, \forall l$ through a dual variable $u \in \mathbb{R}^L$, and decomposes problem~(\ref{jointOpt}) into two subproblems corresponding to the network and physical layers. 

The augmented Lagrangian \newt{for the extended problem with penalty parameter $\rho > 0$ and dual variable $y$}  is
\[
\mathfrak{L}=\min_{n\in\mathcal{\tilde{N}}}\left(r_n\right)-y^T(x-t)-\frac{\rho}{2}\|x - t\|^2_2,
\]
subject to constraints (1a)-(1h), but it is often more convenient to scale the dual variable $y$ and combine the linear and quadratic terms, applying ADMM to
\[
\tilde{\mathfrak{L}}= \min_{n\in\mathcal{\tilde{N}}}\left(r_n\right) - \frac{\rho}{2}\|x - t + u \|^2_2,
\]
where $u=\rho^{-1}y$ and still subject to constraints (1a)-(1h). Both formulations are equivalent, as shown in \cite{boyd11} (section 3.1.1).
The network-layer optimization problem is then:
\begin{subequations}
\begin{align} \label{network}
	&\underset {x \geq 0,(r_n)}{\text{maximize}}
	& &\min_{n\in\mathcal{\tilde{N}}}\left(r_n\right) - \frac{\rho}{2}\|x - t + u \|^2_2 \tag{3}\\
	&\text{subject to}
	& &\sum_l{A_{nl}x_l} = r_n, \; \forall n \in \mathcal{N},  \nonumber 
\end{align}
\end{subequations}
and the physical layer problem is:
\begin{subequations}
\begin{align} \label{physical}
	&\underset {p,w,t \geq 0,(W_g)}{\text{maximize}}
	& &- \frac{\rho}{2}\|x - t + u \|^2_2 \tag{4}\\
	&\text{subject to}
	& &t_l \leq  w_l\log\left( 1+\frac{p_lq_l}{w_lN_0}\right), \; \forall l \in \mathcal{L} \nonumber \\
    &&&(\ref{powerConstraint})-(\ref{individualBwConstraints}).\nonumber
\end{align}
\end{subequations}

Let superscripts denote iteration indices, omitted in the above subproblems for clarity. At iteration $k=1$, the network layer parameters $x^{(1)}$ are obtained from~(\ref{network}) using arbitrary initial values for the dual variables $u^{(0)}$ and auxiliary variables $t^{(0)}$. The physical layer parameters $p^{(1)}$, $w^{(1)}$ and $t^{(1)}$ are then obtained from~(\ref{physical}) using the same $u^{(0)}$ and the obtained values for $x^{(1)}$.  Finally, the dual variable $u^{(1)}$ is updated based on $u^{(0)}$, $t^{(1)}$, and $x^{(1)}$ 
as follows:
\begin{equation}\label{dualUpdate}
u^{(k+1)} = u^{(k)} + x^{(k+1)} - t^{(k+1)}.
\end{equation}
\newt{ADMM has guaranteed convergence} for any $\rho>0$ but in practice the speed of convergence can change depending on its value. Unfortunately, there is no way to know in advance which value will yield the fastest convergence. Convergence results of this algorithm are investigated in Section \ref{numericalExample} through an example.

\subsection{Semi-Distributed Device Optimization Algorithm} \label{3level}

The authors are not aware of any fully distributed algorithm that guarantees convergence to the global optimum using only local coordination among nodes. 
\newt{In any case, such an algorithm would probably require a very large number of iterations, since the constraints in problem~(\ref{jointOpt}) are not local. Information such as data rate from each node needs to reach every potential relay down the chain, then travel back and forth until they converge on a solution}. Instead, a \textit{semi}-distributed algorithm is proposed where each node independently optimizes its own resources based on local channel information and limited communication with a central entity (such as the BS). The algorithm uses ADMM to decompose problem~(\ref{jointOpt}) into an independent subproblem for each node and a shared dual update. One major benefit of the semi-distributed algorithm is that the algorithm can adapt to changing local channel conditions during the iterations of solving the global problem. In the centralized approach, however, the final system parameters are obtained based on the initial channel information sent to the central unit\footnote{The addition energy consumption for message-passing in the semi-distributed approach and the performance gain that can be obtained by updating local channel information are very system dependent and are out of the scope of this work.}.

In the semi-distributed problem there are two sets of coupling variables: the per-link flows $x_l$ and the per-node bandwidths $v_n$. To decouple the per-node optimization subproblems from the global problem, auxiliary sets of variables $t_l, \; \forall l \in \mathcal{L}$ and $b_n, \; \forall n \in \mathcal{\tilde{N}}$ are introduced along with the following equality constraints:
\begin{subequations}
\begin{align}
&t_l = x_l, \; \forall l \in \mathcal{L} \\
&b_n = v_n, \; \forall n \in \mathcal{\tilde{N}}.
\end{align}
\end{subequations}

Even though it appears that the problem complexity increases with the increased number of constraints and variables, it will be shown that such decoupling generates simpler convex subproblems. The set of original variables, $\psi \in \mathbb{R}^{L+N-1}$, and the set of new variables, $z \in \mathbb{R}^{L+N-1}$, are defined as: $\psi = [x_1,...,x_L, v_2,...,v_N]$, $z = [t_1,...,t_L,b_2,...,b_N]$. The two variables $v_1$ and $b_1$, which correspond to the destination node bandwidth, are eliminated.

The variable $\xi \in \mathbb{R}^{L+N-1}$ represents the set of dual variables $\xi = (u, y)$, such that $u \in \mathbb{R}^L$ corresponds to the routing variables and $y \in \mathbb{R}^{N-1}$ corresponds to the bandwidth variables. \newt{The Lagrangian then becomes
 \[
 \tilde{\mathfrak{L}}= \min_{n\in\mathcal{\tilde{N}}}\left(r_n\right) - \frac{\rho}{2}\|\psi - z + \xi \|^2_2,
 \]
 subject to constraints (1a)-(1h).} With this notation and using ADMM, problem~(\ref{jointOpt}) is decomposed into two subproblems. The first subproblem optimizes routing and the amount of bandwidth allocated to each user:
\begin{subequations}
\begin{align} \label{networkAndGroup}
	&\underset {\psi =[x, v] \geq 0,(r_n,W_g)}{\text{maximize}}
	& &\min_{n\in\mathcal{\tilde{N}}}\left(r_n\right) - \frac{\rho}{2}\|\psi - z + \xi \|^2_2 \tag{7}\\
	&\text{subject to}
	& &\sum_l{A_{nl}x_l} = r_n, \; \forall n \in \mathcal{N}  \nonumber \\
	&&&\sum_{n \in \mathcal{G}_g}{v_{n}} \leq W_{g}, \; g \in \left\{1,2,...,M\right\} \nonumber \\
    &&& (\ref{bwConstraint})-(\ref{individualBwConstraints}).\nonumber
\end{align}
\end{subequations}
The second subproblem determines variables $z$, per link power $p_l$, and bandwidth $w_l, \; \forall l$:
\begin{subequations}
\begin{align} \label{nodes3}
	&\underset {p, w, z=[t, b] \geq 0}{\text{maximize}}
	& &- \frac{\rho}{2}\|\psi - z + \xi \|^2_2 \tag{8} \\
	&\text{subject to}
	& &t_l \leq  w_l\log\left( 1+\frac{p_lq_l}{w_lN_0}\right), \; \forall l \in \mathcal{L} \nonumber \\
	&&&\sum_{l \in \mathcal{O}(n)}{w_l} = b_{n},\; \forall n \in \mathcal{\tilde{N}}   \nonumber \\ 
    &&&(\ref{powerConstraint})-(\ref{PmaxConstraint}). \nonumber
\end{align}
\end{subequations}
Subproblems (\ref{networkAndGroup}) and (\ref{nodes3}) are then solved iteratively with dual updates:
\begin{equation}\label{dualUpdates3}
\xi^{(k+1)} = \xi^{(k)} + \psi^{(k+1)} - z^{(k+1)}.
\end{equation}

As demanded, problem~(\ref{nodes3}) can now be decomposed into $N-1$ separate subproblems, such that node $n$ solves the following problem based on its local channel information $q_l, \forall l\in \mathcal{O}(n)$:
\begin{subequations}
\begin{align} \label{physical3}
	&\underset {p_l, w_l, t_l, \forall l \in \mathcal{O}(n),b_n}{\text{maximize}}
	& &- \frac{\rho}{2}\sum_{l \in \mathcal{O}(n)}{\|x_l- t_l + u_l \|^2_2} \tag{10}\\
	&&&- \frac{\rho}{2}\|v_n - b_{n} + y_n \|^2_2 \nonumber\\
	&\text{subject to}
	& &t_l \leq  w_l\log\left( 1+\frac{p_lq_l}{w_lN_0}\right), \; \forall l \in \mathcal{O}(n)\label{capacity3}\\
	&&&\sum_{l\in \mathcal{O}(n)}{p_l\leq P_{\max}}\label{powerConstraint3} \\
	&&& p_l \leq w_l \gamma, \; \forall l \in \mathcal{O}(n) \nonumber \\
	&&&\sum_{l \in \mathcal{O}(n)}{w_l} = b_{n}    \nonumber \\
	&&& p_l, w_l, t_l \geq 0, \; \forall l \in \mathcal{O}(n). \nonumber
\end{align}
\end{subequations}

Furthermore, subproblem~(\ref{networkAndGroup}) is separable into bandwidth $v$ and flow variables $x$. The objective in (\ref{networkAndGroup}) can be rewritten as $\min_{n\in\mathcal{\tilde{N}}}\left(r_n\right) - \frac{\rho}{2}\|x - t + u \|^2_2 - \frac{\rho}{2}\|v - b + y \|^2_2$. Hence this subproblem can further be decomposed into routing and spectrum assignment problems. The routing subproblem is:
\begin{subequations}
\begin{align} \label{network3}
	&\underset {x \geq 0,(r_n)}{\text{maximize}}
	& &\min_{n\in\mathcal{\tilde{N}}}\left(r_n\right) - \frac{\rho}{2}\|x - t + u \|^2_2 \tag{11}\\
	&\text{subject to}
	& &\sum_l{A_{nl}x_l} = r_n, \; \forall n \in \mathcal{N}.  \nonumber 
\end{align}
\end{subequations}
The spectrum assignment subproblem is simply the Euclidean projection onto a constrained set:
\begin{subequations}
\begin{align} \label{group3}
	&\underset {v \geq 0,(W_g)}{\text{maximize}}
	& &- \frac{\rho}{2}\|v - b + y \|^2_2 \tag{12}\\
	&\text{subject to}
	& &\sum_{n \in \mathcal{G}_g}{v_{n}} \leq W_{g}, \; g \in \left\{1,2,...,M\right\} \nonumber \\
    &&& (\ref{bwConstraint})-(\ref{individualBwConstraints}).\nonumber
\end{align}
\end{subequations}
Hence problems (\ref{network3}) and (\ref{group3}) can be solved in parallel and even by two independent entities. The dual variables $u$ and $y$ can also be updated separately as:
\begin{eqnarray}
u^{(k+1)} & = & u^{(k)} + t^{(k+1)} - x^{(k+1)} \label{dualUpdateU}\\
y^{(k+1)} & = & y^{(k)} + b^{(k+1)} - v^{(k+1)}. \label{dualUpdateY}
\end{eqnarray}
%
%
%
The primal residuals corresponding to link flows are $h_1^{(k)} = x^{(k)} - t^{(k)}$ and the primal residuals corresponding to bandwidth are $h_2^{(k)} = v^{(k)} - b^{(k)}$. The dual residuals are then defined as  $s_1^{(k)}= -\rho(t^{(k)} - t^{(k-1)})$ and $s_2^{(k)}= -\rho(b^{(k)} - b^{(k-1)})$.
%

%

Various criteria can be considered as stopping points for the ADMM method. \newt{According to~\cite{boyd11}, one possibility could be:
\begin{eqnarray}\label{stoppingCriteria}
\|h_1^{(k)}\|_2 \leq \epsilon_1^{p},\; \|h_2^{(k)}\|_2 \leq \epsilon_2^{p}, \\
\|s_1^{(k)}\|_2 \leq \epsilon_1^{d},\; \|s_2^{(k)}\|_2 \leq \epsilon_2^{d} \nonumber
\end{eqnarray}
for small predefined $\epsilon_1^p$, $\epsilon_2^p$, $\epsilon_1^d$, and $\epsilon_2^d$ values.}

\subsection{Implementation}

This section focuses on the implementation of the semi-distributed device optimization algorithm introduced above and investigates the required message passing between different entities. All the information regarding channels, network topology, and other such infrastructure can be communicated to the nodes by the central unit (base-station) through control channels. In cellular networks, signaling for link establishment can be done via slight modifications to the standard radio resource control (RRC) connection reconfiguration procedure \cite{ses11}. This procedure can include signaling to the sink nodes of each hop indicating their role as well as providing a list of source nodes that the sink node is providing wireless broadband service to.

Based on the optimization subproblems developed in the previous section, the entity solving problem~(\ref{network3}) (which can be the BS) provides each user $n$ with the flows $x_l,\; \forall l \in \mathcal{O}(n)$ that it should carry on each of its outgoing links and the total amount of bandwidth $v_n$ that it can use. Given $v_n$ and $x_l$, each node calculates $p_l, w_l$, and $t_l,\; \forall l \in \mathcal{O}(n)$ by solving optimization problem~(\ref{physical3}). Each node $n$ then broadcasts $t_l, \; \forall l \in \mathcal{O}(n)$ and $b_n = \sum_{l \in \mathcal{O}(n)}{w_l}$.

\begin{figure}[t]
\centering
\includegraphics[scale =0.3]{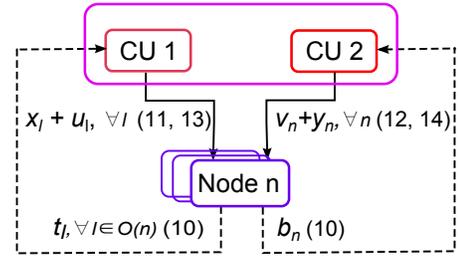}
\caption{\small $3$-level optimization and message passing among different entities.}\label{fig:flowchart}
\end{figure}

The primal variables $x$ and $v$ can be calculated at a single central entity in parallel or by two independent entities. Denote by CU~$1$, the central unit that collects $t_l$ values from all nodes and solves problem~(\ref{network3}) to obtain $x$. \newt{A second central unit (CU~$2$) is in charge of group bandwidths: it} gathers per user calculated bandwidths $b_n$ and solves~(\ref{group3}) to obtain $v$. After each iteration, the dual variables $u$ and $y$ are updated.
The dual variable updates can be performed either at the central units or at each node, resulting in different message passing structures. \newt{Figure~\ref{fig:flowchart} shows the message-passing when the updates are performed at the CU's,} and the corresponding distributed algorithm is summarized in Algorithm~\ref{alg:distributed3}.

Figure~\ref{fig:flowchart} shows the variables that are updated by each entity and the reference number of the problem solved to obtain the corresponding variable. In this configuration, based on initialized values of $u^{(0)}$ and $t^{(0)}$, CU~$1$ solves optimization problem~(\ref{network3}) to obtain $x^{(1)}$. Then the sum of the dual variable and the updated primal variable ($u^{(0)}+x^{(1)}$) is broadcast. Similarly, based on initialized values of $y^{(0)}$ and $b^{(0)}$, CU~$2$ solves optimization problem~(\ref{group3}) to obtain $v^{(1)}$. Then ($y^{(0)} + v^{(1)}$) is broadcast. Each node $n$ captures its required information ($y_n^{(1)} + v_n^{(1)}$) and ($u_l^{(0)} + x_l^{(1)}$), $\forall l \in \mathcal{O}(n)$, \newt{calculates $b_n^{(1)}$ and $t_l^{(1)}, \; \forall l \in \mathcal{O}(n)$ by solving problem~(\ref{physical3}), and broadcasts the updated variables.} During the next iteration, CU~$1$ calculates $u^{(1)}$, and CU~$2$ calculates $y^{(1)}$ based on equations (\ref{dualUpdateU}) and (\ref{dualUpdateY}), respectively.  In this manner, at each iteration, all the dual and primal parameters are updated.
\begin{algorithm}
  \caption{Proposed distributed cross-layer optimization.}
    \label{alg:distributed3}
  \begin{algorithmic}[1]
  		\Statex \textbf{initialize} $u_l$, $y_{n}$, $t_l$, $b_n, \small {\forall l}$ and $\small {\forall n}$
  		\vspace{0.15cm}
  		\Repeat
 				\State perform in parallel:
 				\begin{itemize}
 				\setlength{\itemindent}{1em}
 				\item CU~$1$: given $u_l$ and $t_l$, update $x_l$ (\ref{network3}) and broadcast $x_l + u_l, \small {\forall l}$.
        \item CU~$2$: given $y_n$ and $b_n$, update $v_n$ (\ref{group3}) and broadcast $v_n + y_n, \small {\forall n}$.
        \end{itemize}
        \State Each node $n$: given $x_l + u_l, \; \small{\forall l \in \mathcal{O}(n)},$ $v_n + y_n$ and
        local channel information, update
        \Statex \hspace{0.5cm} $t_l, \small{\forall l \in \mathcal{O}(n)}$ and $b_n$ (\ref{physical3})
        and broadcast results.
        \State CU~$1$ and $2$: update corresponding dual variables $u_l, \small {\forall l}$
        and $y_n, \small {\forall n}$ (\ref{dualUpdates3}).
				\Until{stopping criterion (\ref{stoppingCriteria}) met.}
	\end{algorithmic}
\end{algorithm}

Each update step in Algorithm~\ref{alg:distributed3} is followed by broadcasting the updated parameters. The dual variable update is performed at the CU's, because there are generally more computing resources at these locations as compared to individual nodes. According to problem (\ref{physical3}), the only information that each node requires from the CU's is the sum of $x_l$ and $u_l$ for all its outgoing links and the sum of $v_n$ and $y_n$. Hence in order to reduce the amount of message-passing between the nodes and CU's, the sum of the dual and primal variables $x_l + u_l, \; \forall l$ and $w_n + y_n, \; \forall n$ are broadcast by the CU's instead of the separate variable values (Figure \ref{fig:flowchart}). After convergence, the optimum routing variables $x$ and physical-layer resources $p$ and $w$ are obtained.

In summary, the semi-distributed optimization algorithm works as follows: Each node $n$ gathers and stores the values of $x_l + u_l,\; \forall l \in \mathcal{O}(n)$, $v_n + y_n$. Then it performs an Euclidean projection onto a set of constraints (\ref{physical3}), which has very small complexity, \newt{obtaining} $t_l,\; \forall l \in \mathcal{O}(n)$, and $b_n$. Each node then broadcasts its updated results. Two central units, CU~$1$ and CU~$2$, that can be the same or independent entities, gather the updated values of $t_l$ and $b_n$, respectively. CU~$1$ and CU~$2$ then solve (\ref{network3}) and (\ref{group3}), respectively, in parallel. These problems also have very small complexity. The results of such optimizations are $x_l, \; \forall l \in \mathcal{L}$, and $v_n, \; \forall n \in \mathcal{\tilde{N}}$. CU~$1$ and CU~$2$ then update the dual variables $u$ and $y$, respectively, while storing a local copy of each variable for the next dual variable update.

In theory, the proof of convergence for ADMM requires that all nodes update $t_l, \; \forall l \in \mathcal{O}(n)$ and $b_n$ in step-3 before the algorithm can proceed to step-4. In most practical cases, however, the algorithm still converges even if some nodes occasionally fail to perform the update. A good practical approach is to allow partial updates (only a subset of nodes update their local variables) during the first few iterations, and impose that all nodes perform the update in subsequent ones. This approach guarantees convergence to the same solution, while allowing faster execution of the algorithm in cases where some nodes are significantly slower than others. Wei and Ozdaglar proposed in \cite{wei13} an asynchronous distributed version of ADMM that converges in $O(1/k)$, but it assumes that all entities send updates infinitely often and only guarantees convergence with probability 1, unlike regular ADMM. The number of iterations required with this asynchronous version can also be significantly larger than with regular ADMM.

The complexity of Algorithm~\ref{alg:distributed3} scales gracefully to large number of nodes. Step-2 consists of solving a simple quadratic program (\ref{network3}) and performing a Euclidean projection onto the convex set specified in (\ref{group3}). Both of these can be easily scaled to large number of variables. Step-3 is performed independently by each node, so its complexity does not grow. Finally, Step-4 computes a linear combination of optimized variables, which again is a very simple operation whose complexity increases linearly with the number of variables. The amount of data broadcast by the nodes also increases linearly with the number of nodes and links. In low-power large scale networks such as wireless sensor networks (WSN), the message-passing overhead can be comparable to the actual data size. The proposed algorithm can then demolish the low-power objective of WSNs; however, it can prove beneficial in dense areas populated with devices that have higher energy resources.

\subsection{A Numerical Example} \label{numericalExample}
\begin{figure}[t]
\centering
\includegraphics[scale =0.45]{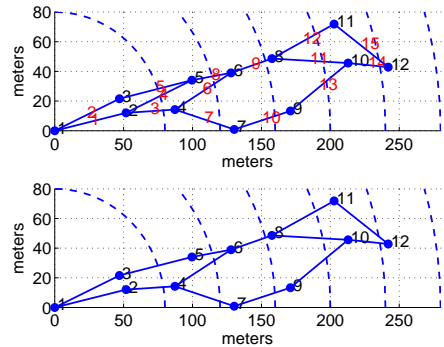}
\caption{\small Top: simple node configuration (N = 12), bottom: resulting optimum routing configuration.}\label{fig:simpleNodeConfig}
\end{figure}

Figure~\ref{fig:simpleNodeConfig} illustrates the algorithm operation over a circular cellular area of radius $280$m. A total of $11$ nodes (with destination $N = 12$) are randomly distributed in this area, creating $6$ groups separated by $d = 40$m (except $d_1 = 2d$). We assume that a total bandwidth of $W_{\max} = 10$MHz is available with a carrier frequency of $800$Mhz. The receiver noise power spectral density is $N_0 = 10^{-11}$W/MHz. The reference distance $l_0$ in constraint~(\ref{powerConstraint}) is set to $1$ and the pathloss exponent is $a = 4$. Each user's maximum transmission power is set to $P_{\max} = 0.5$W.

\begin{figure}[t]
\centering
\includegraphics[scale =0.4]{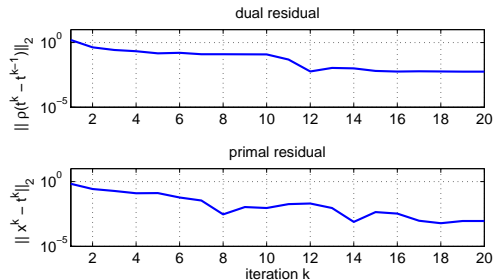}
\caption{\small Norms of primal and dual residuals versus iteration for layered optimization problem.}\label{fig:residuals2}
\end{figure}

\begin{figure}[t]
\centering
\includegraphics[scale =0.4]{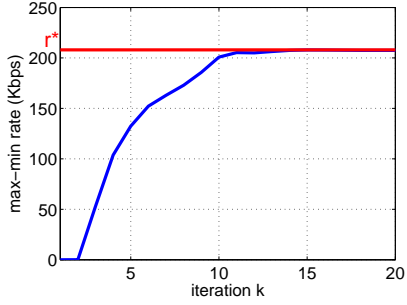}
\caption{\small Max-min rate convergence for layered optimization problem.}\label{fig:rateConvergence2}
\end{figure}

 Setting the angular threshold $\theta = 10^o$ and the distance threshold $d_{th} = 1.5d$ in the link assignment step, a total of $L = 15$ initial links are generated, as shown in the top plot of Figure~\ref{fig:simpleNodeConfig}. CVX, a package for specifying and solving convex programs \cite{cvx}, is used to solve the centralized optimization problem (\ref{jointOpt}), resulting in a max-min rate of $208$Kbps. For the disjoint network layer and physical layer optimization problems, the penalty parameter is set to $\rho = 1$ (for variations refer to \cite[Ch 3.4]{boyd11}) and routing parameters $t^{(0)}$ are initialized to $1$. Solving the individual subproblems~(\ref{network})~and~(\ref{physical}) iteratively via ADMM results in the same max-min rate of $208$Kbps, as shown in Figure~\ref{fig:rateConvergence2}. The norms of the primal and dual residuals at each iteration are depicted in Figure~\ref{fig:residuals2}. The results show that \newt{the centralized algorithm} converges in about $15$ iterations.

\begin{figure}[t]
\centering
\includegraphics[scale =0.3]{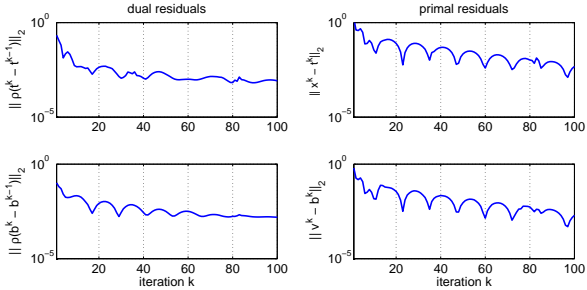}
\caption{\small Norms of primal and dual residuals versus iteration for the semi-distributed optimization problem.}\label{fig:residuals}
\end{figure}

\begin{figure}[t]
\centering
\includegraphics[scale =0.31]{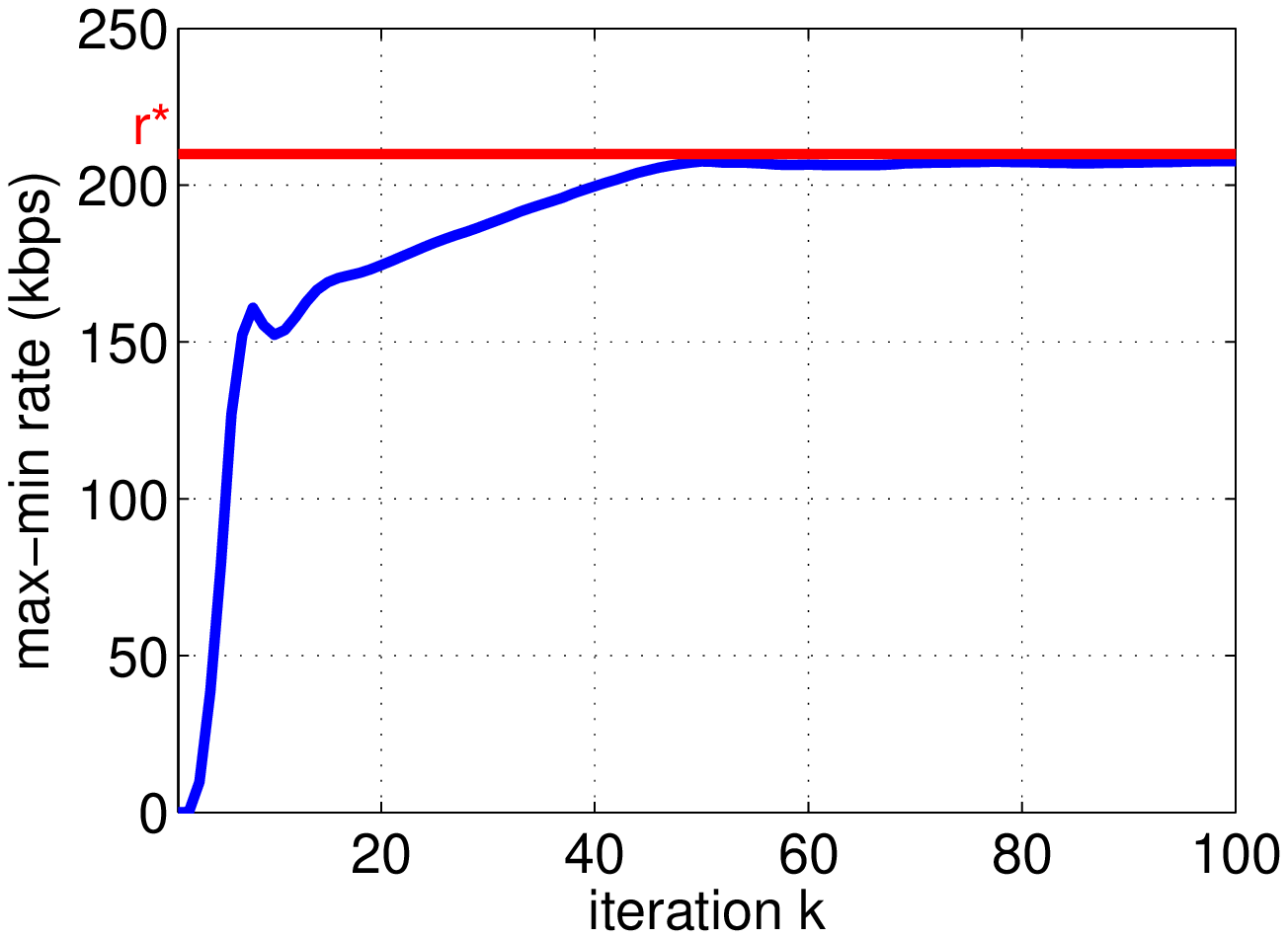} \includegraphics[scale =0.29]{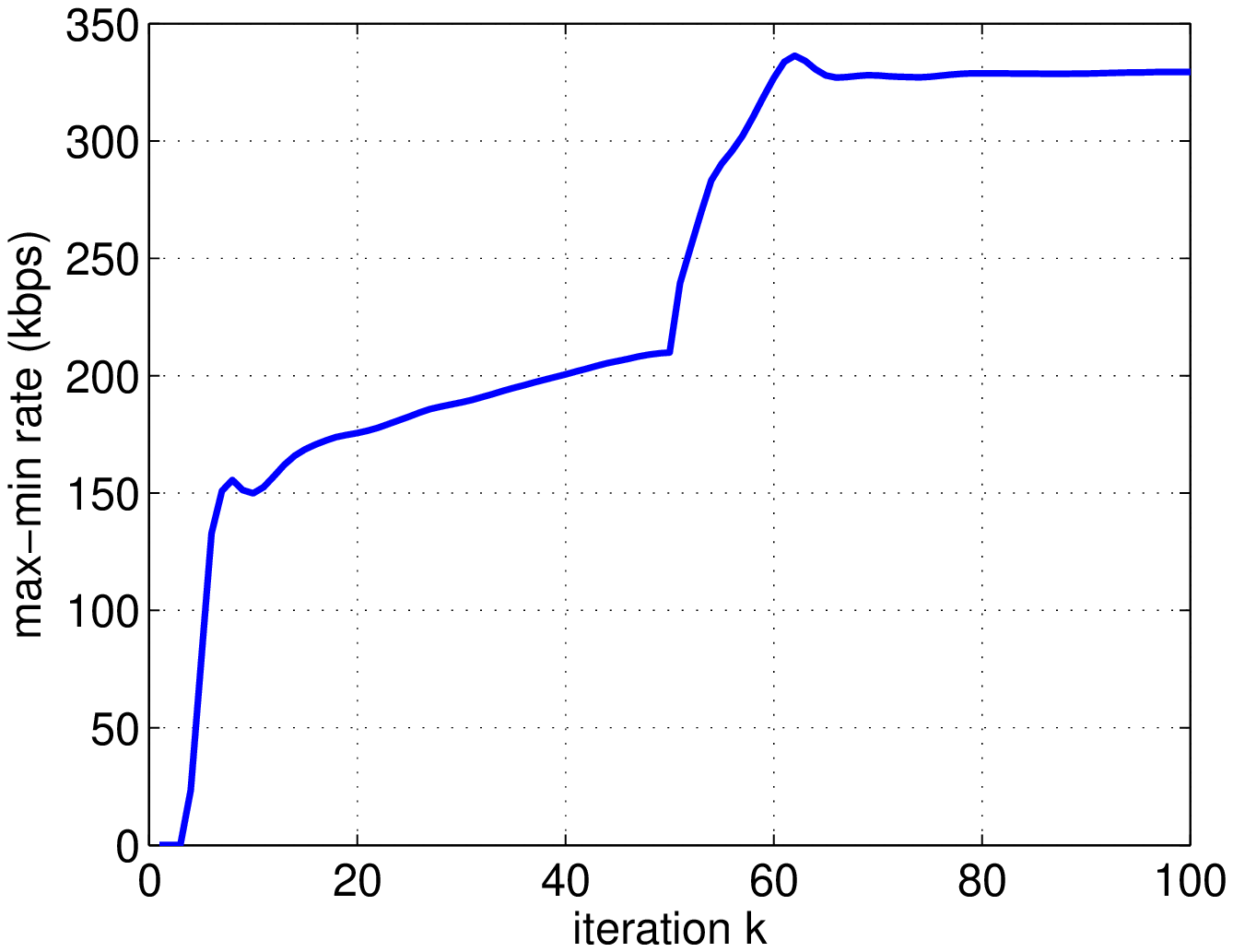}
\caption{\small Max-min rate convergence for the semi-distributed optimization problem for a time-invariant channel (left) and $N_0$ changed after $50$ iterations (right).}\label{fig:rateConvergence3}
\end{figure}

Before implementing the subproblems (\ref{physical3}), (\ref{network3}), and (\ref{group3}) of the semi-distributed algorithm \newt{in Section~\ref{3level}}, it is important to investigate their nature. Optimization problem~(\ref{physical3}) is performed by all the nodes (in parallel) and has a quadratic objective and linear constraints other than constraint~(\ref{capacity3}). By approximating the $\log$ function with a piecewise linear function, problem~(\ref{physical3}) can be converted to a simple Quadratic Programming (QP) problem with a relatively small number of variables. Then CVXGEN \cite{mat10}, which can generate fast custom code for small QP-representable convex optimization problems, can be used to solve the problem.

A sequential convex programming approach \cite{ima08} is used to convert subproblem~(\ref{physical3}) into a QP: the capacity term in (\ref{capacity3}), $c(w_l, p_l) = w_l\log(1+\frac{p_lq_l}{w_lN_0})$, is approximated by a piece-wise linear function: $\tilde{c}(\beta) = \min_{(a_1, a_2) \in \Sigma} \left\{a_1^T \beta + a_2\right\}$, where 
$\beta = \left(p_l, w_l\right)$ and $\Sigma$ represents a set of planes tangent to $c(\beta)$. For simplicity, $\Sigma$ in this case consists of only $10$ planes. Consequently, constraint (\ref{capacity3}) is replaced with the following linear constraint:
\begin{equation*}
t_l \leq  \min_{(a_1, a_2) \in \Sigma} \left\{a_1^T\beta+ a_2\right\}, \; \forall l \in \mathcal{O}(n)
\end{equation*}

This approximation can be repeatedly re-adjusted during the execution of the algorithm to choose planes that are tangent at the current iterate, and hence the approximation error is progressively reduced. Eventually, when a point is reached where planes are tangent at the iterate points and the iterates remain constant (reached an equilibrium point), it is ensured that the optimal solution with zero approximation error is achieved.

The number of variables in this subproblem depends on the number of outgoing links from each node. Each node's unknown parameters are $t_l$, $p_l$ and $w_l$, and hence if node $n$ has $L_n$ outgoing links, it will have $3L_n$ variables. Since the nodes in Figure~\ref{fig:simpleNodeConfig} only have $1$ or $2$ outgoing links, each node has a $3$ or $6$ variable QP problem to solve.

Subproblem (\ref{group3}) is readily in QP format and subproblem (\ref{network3}) can be easily converted to a QP by replacing it with its epigraph form. The problem then becomes
\begin{subequations}
\begin{align} \label{network3}
	&\underset {x \geq 0,(\nu)}{\text{maximize}}
	& &\nu - \frac{\rho}{2}\|x - t + u \|^2_2 \tag{11}\\
	&\text{subject to}
	& &\sum_l{A_{nl}x_l} = \nu, \; \forall n \in \mathcal{\tilde{N}},  \nonumber 
\end{align}
\end{subequations}
 Both subproblems (\ref{network3}) and (\ref{group3}) with $L = 15$ and $N = 12$ variables, respectively, can now be implemented using CVXGEN, as well.

%
%

 Setting $\rho=0.5$ and initializing the flows $t^{(0)}$ and bandwidths $w^{(0)}$ to $1$, a max-min rate of $208$Kbps is obtained through the above approach. The primal and dual residual norms at each iteration are represented in Figure~\ref{fig:residuals}, and Figure~\ref{fig:rateConvergence3} (left) represents the max-min rate convergence results. Simulations using CVXGEN (on an Intel core i7, 2.7 GHz processor) \newt{took an average of $0.327$ms to solve problem~(\ref{network3}) and $0.132$ms to solve problem~(\ref{group3}). Problem~(\ref{physical3}), solved in parallel by all nodes, took $0.145$ms in average. In this scenario, the semi-distributed} algorithm requires approximately $50$ iterations to converge. Ignoring the time required for message passing $50 \times \left(0.145 + \max\left\{0.132, 0.327\right\}\right) = 23.6$ms are required on average to obtain the optimal routing and resource allocation solution. Ignoring the links that have negligible flows (e.g. less than $10$Kbps in this case), the bottom plot of Figure~\ref{fig:simpleNodeConfig} is obtained, where Link $4$ has been eliminated due to carrying only $8$Kbps. If desired, the resource allocation can then be re-optimized based on the updated links.

 The number of iterations and time required for the algorithm to converge can be reduced significantly through ``warm start'' initialization techniques \cite[ch.4.3]{boyd11}. The rapid convergence of ADMM when the initial point is close to the optimal solution also allows the semi-distributed implementation to adapt to changing network conditions. Figure~\ref{fig:rateConvergence3} (right) shows the convergence of the max-min rate when each node's receiver noise power spectral density $N_0$ is \newt{scaled} by a random factor between $\frac{1}{2}$ and $\frac{5}{2}$ after the first 50 iterations. The nodes do not need to notify the CUs of this change, they just use the new value when solving their subproblems. ADMM took 50 iterations to converge from the initial conditions, but only 12 more iterations to adapt to the new values for $N_0$.



\section{Performance Evaluation}\label{sec:performance}

This section evaluates the performance of the proposed hierarchical node configuration and cross-layer optimization algorithms by \newt{using the direct communication in (\ref{DM}) as the base line for comparison}. Simulations consider a single sector of a cellular network. However, the results can be easily extended to the entire cellular area, allowing spectrum to be reused among different sectors as well as among the groups within a sector.

\subsection{Simulation Setup}
\begin{figure}[t]
\centering
\includegraphics[scale =0.45]{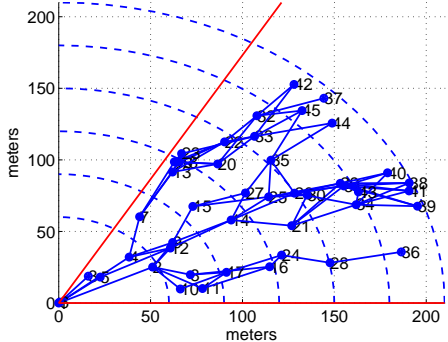}
\caption{\small Nodes distributed randomly ($N = 45$) with initial set of possible links ($L=86$).}\label{fig:nodeConfiguration}
\end{figure}

\newt{An instance of the simulation scenario is shown in Figure~\ref{fig:nodeConfiguration}. It consists of }a single sector from a $6$-sector circular cellular network with radius $210$ meters. A total of $44$ users ($N = 45$ with destination node) are randomly distributed within this sector and the common destination node is located at $(0,0)$. The sector is sub-divided such that the radial distance between consecutive groups is $d = 30$m.  Since all nodes within group~$1$ communicate directly with the node at $(0,0)$, the radial distance to the boundary of group $1$ ($d_1$) is set to double the regular distance $d$. Such value keeps the average link lengths within group $1$ similar to other links in the network. \newt{According to the routing-link assignment policy, with a distance threshold of $1.5 d$ and angular threshold of $15^0$, a total of $86$ links were generated for the network in Figure~\ref{fig:nodeConfiguration}.}

The number of hops can be adjusted by varying $d$. For low rate requirements, as the number of hops increases, the total required transmission power from a source to destination decreases. However, there is a limit on the number of hops, since there must be at least one node per group \newt{to take advantage of the frequency reuse}. \newt{Furthermore, the reduction in transmission power with increasing number of hops} comes at the expense of increasing the number of nodes cooperating to transmit data and a larger routing table. 

\newt{The total available bandwidth was assumed to be} $W_{\max} = 10$MHz with carrier frequency of $800$Mhz. The receiver power spectral density is $N_0 = 10^{-11}$W per $1$MHz bandwidth. The reference distance $l_0$ in constraint (\ref{powerConstraint}) is set to $1$ and the pathloss exponent $a = 4$. Four different methods are compared in terms of their average performance: multi-hop configurations with frequency reuse of $f = 3$ and $f =4$, multi-hop configuration without reuse \newt{($f=\infty$)}, and direct communication (\ref{DM}). \newt{The results were averaged over $1000$ randomly generated networks, constrained to have at least one node in each group.}  A negative factor of power sum, $-\epsilon \sum{p_l}$, with $\epsilon = 10^{-6}$ is used in the simulations, as described in Section~\ref{sec:algorithm}.

\subsection{Results and Discussion}

\begin{figure}[t]
\centering
\includegraphics[scale =0.45]{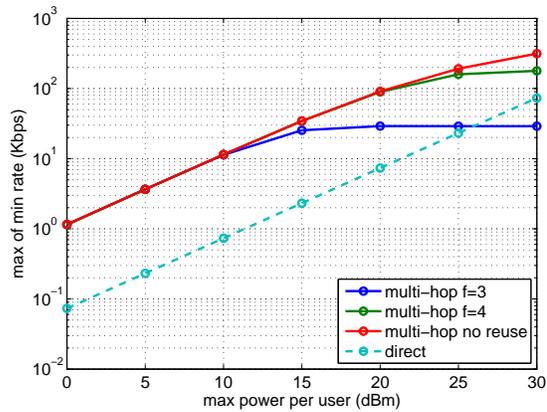}
\caption{\small Per user rate vs. per user maximum transmission power.}\label{fig:rateVsPmax}
\end{figure}

\begin{figure}[t]
\centering
\includegraphics[scale =0.45]{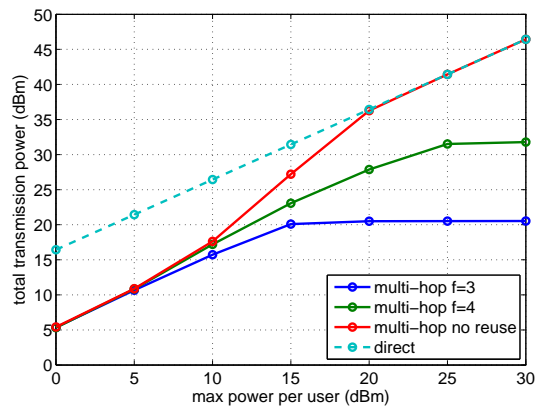}
\caption{\small  Total network transmission power vs. per user maximum transmission power ($N = 45$).}\label{fig:powerVsPmax}
\end{figure}

 The maximum transmission power per device $P_{\max}$ is varied between $0$dBm  and $30$dBm and the minimum rate among all users is plotted in Figure~\ref{fig:rateVsPmax}. Furthermore, the total transmission power employed by the network for each $P_{\max}$ is represented in Figure~\ref{fig:powerVsPmax}. It is evident from these figures that the performance of multi-hop configurations with reuse can be significantly limited because of transmission power limits that avoid interference. To solve convex problems ({\ref{jointOpt}) and (\ref{DM}), CVX, a package for specifying and solving convex programs \cite{cvx} is used.

 \newt{The minimum rate and total power varied significantly among the $1000$ networks that were averaged, but the relative performance of each scheme remained fairly constant. For example, when the maximum power per user is limited to $P_{\max}=20$dBm, the standard deviation in the minimum transmission rate for the multi-hop scheme was $8.6$ Kbps when $f=3$, $45$ Kbps when $f=4$, and $47$ Kbps with no reuse, but the rates with the latter two were nearly identical and better than with the $f=3$, for all networks. Similarly, the total transmission power had a standard deviation of $0.5$ dBm for $f=3$, $2$ dBm for $f=4$, and $1.8$ dBm without frequency reuse, but the relative positions of the curves remained constant. The standard deviations with the direct transmission scheme were negligible in all cases.}

Problem~(\ref{jointOpt}) contains two constraints on transmission power: constraint (\ref{powerConstraint}), which limits the transmission power on each link and constraint (\ref{PmaxConstraint}), which limits per node transmission power $P_{\max}$ (i.e., the total power a node transmits over all its outgoing links). This implies that when $P_{\max}$ is large, the transmission powers are limited by constraint (\ref{powerConstraint}) and when $P_{\max}$ is small, constraint (\ref{PmaxConstraint}) dominates. Figure \ref{fig:powerVsPmax} shows the $P_{\max}$ threshold values for frequency reuse factors of $f =3,$ ($15$dBm) and $f=4$ ($25$dBm), after which increasing $P_{\max}$ does not change the total network transmission power and hence does not increase the achievable rate. Figure~\ref{fig:powerVsPmax} also shows that total network transmission power increases linearly with $P_{\max}$ in direct mode where constraint~(\ref{powerConstraint}) does not apply.

At maximum transmission power of $0$dBm, as Figure~\ref{fig:rateVsPmax} depicts, multi-hopping provides a factor of $10$ increase in data rate relative to direct communication, while approximately $10$ times less transmission power is required (Figure~\ref{fig:powerVsPmax}). However, this performance gain decreases as $P_{\max}$ increases. As theoretical studies of spatial reuse and multi-hopping \cite{kum07} suggest, 
using direct communication between source and destination is preferred over spatial reuse and multi-hopping when large transmission powers are allowed. However, for limited power sources, multi-hopping and spatial reuse perform better than direct mode.

At moderately low transmission powers the configurations with reuse factors of $3$ and $4$ obtain the same data rate as that without reuse, but the total network transmission power for each case is different. In general, for the same data rate, the required transmission power increases with the reuse factor. When $P_{\max}=10$dBm, the total transmission power required by the multi-hop scheme without reuse is $2.7$ times the amount of power required when $f=3$ and $1.5$ times the amount of power required in the $f=4$ case, despite all three provide the same minimum rate.

The above results, obtained through joint optimization of routing and resource allocation, often require data to flow from a source node through multiple routes to the destination node. The number of used routes is determined in part by the initial pool of links selected during the link assignment step of Section~\ref{subsec:linkAssignment}. One method for reducing the number of used routes, albeit with a loss in optimality, is to restrict the number of outgoing links from each node by reducing $d_{th}$ and $\theta$ .

At the expense of further reducing network performance, a less complex approach is to enforce a single route from each source to the destination. This can be done by eliminating routing from the joint optimization framework and simply allowing each node to connect to the closest neighbor in the group above it. Such approach even though simpler to implement, performs significantly worse than cross-layer optimization, because the nodes that are located near the boundary of the groups will be congested with incoming traffic. \newt{Another alternative would be to solve the problem with multiple routes and have each node use all its resources for the most promising outgoing link. The performance loss with each of these approaches }depends on many network parameters such as the number of nodes present and their configuration.

\section{Summary}\label{sec:summary}

The demand for better spectrum use grows dramatically as the wireless technology and mobile devices progress rapidly. This demand is more stressed in areas where radio devices are densely packed and need to communicate with a common destination node. This paper proposes employing these radio devices as both relays and computing resources for better system performance. In this method, each device is used to relay other user's data and perform parallel optimization of global resource allocation. Simulation results show that the proposed method can boost each user's rate by a factor of $10$ while using lower transmission power. However this gain is at the expense of higher system complexity.

It is shown that multi-hop communication with reuse has better performance when transmission power is limited by device power and not interference. By further increasing spectrum reuse, the available bandwidth increases, but at the expense of tightening transmission power limits to mitigate co-channel interference. Pathloss can be further reduced and transmission power lowered by increasing the number of hops between each node and the destination, but again at the expense of higher system complexity with larger number of hops.


\section*{Acknowledgment}
The authors would like to thank Professor Stephen Boyd for his valuable discussions. This paper was funded by the Deanship of Scientific Research (DSR), King Abdulaziz University, under grant No. (11-15-1432 HiCi). The authors, therefore, acknowledge with thanks DSR technical and financial support.

\bibliographystyle{./IEEEtran}
\bibliography{./IEEEabrv,./SpatialReuse}

\begin{thebibliography}{10}
\providecommand{\url}[1]{#1}
\csname url@samestyle\endcsname
\providecommand{\newblock}{\relax}
\providecommand{\bibinfo}[2]{#2}
\providecommand{\BIBentrySTDinterwordspacing}{\spaceskip=0pt\relax}
\providecommand{\BIBentryALTinterwordstretchfactor}{4}
\providecommand{\BIBentryALTinterwordspacing}{\spaceskip=\fontdimen2\font plus
\BIBentryALTinterwordstretchfactor\fontdimen3\font minus
  \fontdimen4\font\relax}
\providecommand{\BIBforeignlanguage}[2]{{%
\expandafter\ifx\csname l@#1\endcsname\relax
\typeout{** WARNING: IEEEtran.bst: No hyphenation pattern has been}%
\typeout{** loaded for the language `#1'. Using the pattern for}%
\typeout{** the default language instead.}%
\else
\language=\csname l@#1\endcsname
\fi
#2}}
\providecommand{\BIBdecl}{\relax}
\BIBdecl

\bibitem{lin06}
X.~Lin, N.~Shroff, and R.~Srikant, ``A tutorial on cross-layer optimization in
  wireless networks,'' \emph{Selected Areas in Communications, IEEE Journal
  on}, vol.~24, no.~8, pp. 1452--1463, 2006.

\bibitem{goldsmith}
A.~Goldsmith, \emph{Wireless Communications}.\hskip 1em plus 0.5em minus
  0.4em\relax New York, NY, USA: Cambridge University Press, 2005.

\bibitem{neu05}
C.~D. A.~S. Michael~Neufeld, Jeff~Fifield and D.~Grunwald, ``Softmac - flexible
  wireless research platform,'' \emph{Fourth Workshop on Hot Topics in Networks
  (HotNets-IV)}, 2005.

\bibitem{san03}
T.~S.~R. Shakkottai, Sanjay and P.~C. Karlsson, ``Cross-layer design for
  wireless networks.'' \emph{Communications Magazine, IEEE 41.10}, pp. 74--80,
  2003.

\bibitem{vin05}
V.~Srivastava and M.~Motani., ``Cross-layer design: a survey and the road
  ahead.'' \emph{Communications Magazine, IEEE 43.12}, pp. 112--119, 2005.

\bibitem{boyd04}
L.~Xiao, M.~Johansson, and S.~Boyd, ``Simultaneous routing and resource
  allocation via dual decomposition,'' \emph{Communications, IEEE Transactions
  on}, vol.~52, no.~7, pp. 1136 -- 1144, july 2004.

\bibitem{boyd03}
M.~Johansson, L.~Xiao, and S.~Boyd, ``Simultaneous routing and power allocation
  in cdma wireless data networks,'' vol.~1, pp. 51--55 vol.1, 2003.

\bibitem{tim11}
B.~Timus, P.~Soldati, D.~Kim, and J.~Zander, ``Cross-layer resource allocation
  model for cellular-relaying network performance evaluation,'' \emph{Vehicular
  Technology, IEEE Transactions on}, vol.~60, no.~6, pp. 2765--2776, 2011.

\bibitem{yu07}
T.~Ng and W.~Yu, ``Joint optimization of relay strategies and resource
  allocations in cooperative cellular networks,'' \emph{Selected Areas in
  Communications, IEEE Journal on}, vol.~25, no.~2, pp. 328--339, 2007.

\bibitem{BB}
\BIBentryALTinterwordspacing
S.~Boyd. (2011) Branch and bound methods. [Online]. Available:
  \url{http://stanford.edu/class/ee364b/lectures/bb\_slides.pdf}
\BIBentrySTDinterwordspacing

\bibitem{boyd11}
S.~Boyd, N.~Parikh, E.~Chu, B.~Peleato, and J.~Eckstein, ``Distributed
  optimization and statistical learning via the alternating direction method of
  multipliers,'' \emph{Foundations and Trends in Machine Learning}, vol.~3,
  no.~1, pp. 1--122, 2011.

\bibitem{mung07}
D.~Palomar and M.~Chiang, ``Alternative distributed algorithms for network
  utility maximization: Framework and applications,'' \emph{Automatic Control,
  IEEE Transactions on}, vol.~52, no.~12, pp. 2254--2269, 2007.

\bibitem{jia12}
J.~Liu and H.~D. Sherali, ``A distributed newton's method for joint multi-hop
  routing and flow control: Theory and algorithm.'' \emph{INFOCOM, 2012
  Proceedings IEEE}, pp. 2489 -- 2497, 2012.

\bibitem{shen12}
C.~Shen, T.-H. Chang, K.-Y. Wang, Z.~Qiu, and C.-Y. Chi, ``Distributed robust
  multicell coordinated beamforming with imperfect csi: An admm approach,''
  \emph{Signal Processing, IEEE Transactions on}, vol.~60, no.~6, pp.
  2988--3003, 2012.

\bibitem{mot12}
J.~Mota, J.~Xavier, P.~Aguiar, and M.~Puschel, ``Distributed admm for model
  predictive control and congestion control,'' in \emph{Decision and Control
  (CDC), 2012 IEEE 51st Annual Conference on}, 2012, pp. 5110--5115.

\bibitem{berts}
D.~P. Bertsekas, \emph{Network Optimization: Continuous and Discrete
  Models}.\hskip 1em plus 0.5em minus 0.4em\relax Belmont, MA, USA: Athena
  Scientific, 1998.

\bibitem{san10}
M.~R. Sanatkar and A.~Mohammadi, ``Scalability analysis of wireless sensor
  networks using analytical techniques,'' \emph{8th Annual Communication
  Networks and Services Research Conference}, vol.~2, no.~1, pp. 298--303,
  2010.

\bibitem{boyd}
S.~Boyd and L.~Vandenberghe, \emph{Convex Optimization}.\hskip 1em plus 0.5em
  minus 0.4em\relax New York, NY, USA: Cambridge University Press, 2004.

\bibitem{ned09}
A.~Nedic and A.~Ozdaglar, ``Distributed subgradient methods for multi-agent
  optimization,'' \emph{Automatic Control, IEEE Transactions on}, vol.~54,
  no.~1, pp. 48--61, 2009.

\bibitem{ses11}
S.~Sesia and et~al, \emph{LTE, The UMTS Long Term Evolution: From Theory to
  Practice}.\hskip 1em plus 0.5em minus 0.4em\relax Wiley and sons: 2nd ed.,
  2011.

\bibitem{wei13}
E.~Wei and A.~Ozdaglar, ``On the $o(1/k)$ convergence of asynchronous
  distributed alternating direction method of multipliers.'' \emph{arXiv
  preprint arXiv:1307.8254}, 2013.

\bibitem{cvx}
I.~CVX~Research, ``{CVX}: Matlab software for disciplined convex programming,
  v2.0 beta,'' \url{http://cvxr.com/cvx}, Sep. 2012.

\bibitem{mat10}
J.~Mattingley and S.~Boyd, ``Cvxgen: A code generator for embedded convex
  optimization,'' \emph{Optimization and Engineering}, vol.~13, no.~1, pp.
  1--27, 2010.

\bibitem{ima08}
A.~Imamoto and B.~Tang, ``A recursive descent algorithm for finding the optimal
  minimax piecewise linear approximation of convex functions,'' in \emph{World
  Congress on Engineering and Computer Science.}\hskip 1em plus 0.5em minus
  0.4em\relax IEEE, 2008, pp. 287--293.

\bibitem{kum07}
V.~Ramaiyan and A.~Kumar, ``On the limits of spatial reuse and cooperative
  communication for dense wireless networks,'' \emph{Information Theory for
  Wireless Networks, 2007 IEEE Information Theory Workshop on}, pp. 1--5, July.

\end{thebibliography}

\end{document}